\documentclass[pre,superscriptaddress,showpacs,
               twocolumn,amsfonts,amsmath]{revtex4}
\usepackage[final]{graphicx}
\newlength\figurewidth
\AtBeginDocument{\setlength\figurewidth{.9\linewidth}} 

\let\bs\boldsymbol
\def\kB{k_{\text{B}}}
\def\kT{\kB T}
\def\f#1{\vec {f^{\vphantom{\text{#1}}}}{\vphantom{f}}^{\text{#1}}}
\def\fr{\f{r}}
\def\fd{\f{d}}
\def\fp{\f{p}}
\def\fe{\f{ext}}

\DeclareMathOperator\erfc{erfc}
\DeclareMathOperator\LT{LT}

\usepackage{color}

\begin{document}

\title{Event-Driven Brownian Dynamics for Hard Spheres}

\date\today

\def\edin{%
  \affiliation{Scottish Universities Physics Alliance,
  School of Physics, The University of Edinburgh,
  JCMB King's Buildings, Edinburgh EH9 3JZ, U.K.}}
\def\romasoft{%
\affiliation{Dipartimento di Fisica and INFM-CRS SOFT, Universit\'a di Roma ``La Sapienza'', P.le Aldo Moro 2, 00185 Roma, Italy }}
\def\romasmc{%
\affiliation{Dipartimento di Fisica and INFM-CRS SMC, Universit\'a di Roma ``La Sapienza'', P.le Aldo Moro 2, 00185 Roma, Italy }}
\author{A.~Scala}\romasmc
\author{Th.~Voigtmann}\edin
\author{C.~De~Michele}\romasoft


\begin{abstract}
Brownian dynamics algorithms integrate numerically Langevin equations and
allow to probe long time scales in simulations. A common requirement for such
algorithms is that interactions in the system should vary little during an
integration time step: therefore, computational efficiency worsens as the
interactions become steeper. In the extreme case of hard-body interactions,
standard numerical integrators become ill defined. Several approximate schemes have been invented to handle such cases with little emphasis on testing the correctness of the integration scheme. Starting from the two-body Smoluchowsky equation, we discuss a general method for the overdamped Brownian dynamics 
of hard-spheres, recently developed by one of us. We test the accuracy of the algorithm with the exact solution of the Smoluchowsky equation in the case of two body collisions and in the low-density limit.
\end{abstract}


\maketitle


\section{Introduction}

The simulation of interacting Brownian particles, called Brownian Dynamics (BD) simulation \cite{Ermak}, has become an important tool in condensed matter, colloidal and biological physics. In particular at high densities, the excluded volume has a major effect on the dynamics of the Brownian particles. Consequently, many model colloids are well described by effective pair interactions that are hard-sphere (HS) like, or have a HS core with additional potential tails.
This places the hard-sphere system (with or without random `Brownian' forces)
among the most important reference systems for the theories in the field \cite{HansenMcDonaldBook,PuseyChapter}. However, such theories are in
general only approximate, requiring extensive testing through experiment
and computer simulation. Yet, the simulation of hard spheres with Brownian
dynamics is less than straightforward, because of the singular nature of
the interaction potential: most methods dealing with the numerical integration of stochastic differential equations (SDEs) require a certain degree of smoothness in all the interactions.

Up to now, several hard-core BD algorithms have been proposed and applied in the past \cite{Cichocki90,Heyes93,Moriguchi95,Sillescu94,Foss2000,Strating99,Tao06}, with a varying degree of justification. Our aim here is to review these approaches and to compare them with a novel approach developed by one of us \cite{Foffi05}, called De Michele's algorithm in the following.
We investigate the convergence to the true solutions of the singular SDE by studying special situations where the theoretical behavior is still under control, in order to validate the correctness of the approach.
This is the first step in a programme to extend De Michele's algorithm to more complicated cases, such as taking into account inertial effects (assumed to be negligible in proper BD), or more complicated singular interactions.

Of course the hard-sphere potential ($\beta V(r)=\infty$ if two particles overlap, zero otherwise, where $\beta=1/(\kT)$ is the inverse temperature), is a purely theoretical concept. One may view steeply repulsive `soft-sphere' potentials, e.g., $V(r)\propto r^{-n}$ with large $n$, as more convenient model systems for colloids \cite{Heyes93,Guevara03}, and even try to infer HS ($n\to\infty$) behavior by mapping the $n$-dependent results, taking into account structural information about the system \cite{Guevara03,FuchsPrivateComm}. Such soft-sphere methods are of course hampered by having to use smaller and smaller integration time-steps for increasing $n$ and careful extrapolation to $n\to\infty$. Therefore, a
proper HS-BD algorithm is worthwhile.

We will not discuss hydrodynamic interactions (HI), i.e., the solvent-induced interactions that are present in typical experimental realizations of Brownian systems. Taking them into account properly ensures that hard spheres cannot overlap, due to divergent lubrication forces \cite{Brady93}. To deal with HI, several computational methods have been developed, for example Stokesian Dynamics \cite{Brady88},
Lattice Boltzmann simulations \cite{Ladd94,Ladd01,Rjoy04,Rjoy05}, Dissipative Particle Dynamics \cite{Hoogerbrugge92,Groot97}, or Fluid Particle methods \cite{Tanaka00}. Usually they either deal with softened interactions again and/or are rather time-consuming: depending on the method, a huge number of degrees of freedom needs to be considered, or the intricate nature of the non-pairwise-additive long-range HI forces the use of elaborate schemes. While the theory of HI is rather well understood at low particle densities, much less is known at high densities, and theories often proceed by claiming them irrelevant. A non-HI simulation therefore still has its place in testing such theories, and in circumventing the huge effort of the HI methods, should the claim be true.

Testing the goodness of the approximations inherent to all existing HS-BD
algorithms has up to now received little attention. One needs to test properties
that are inherent both to the Brownian dynamics of the system and to the
hard-core collisions. Testing for diffusive behavior is obviously not enough,
in as far as at high densities and long times it is related to the chaotic nature of the many-body problem and not to the details of the implemented equation of motion or its correctness. In some previous work, static quantities like the
radial distribution function $g(r)$ have been checked \cite{Strating99},
but such comparisons do not test whether an HS-BD algorithm properly
discretizes the Langevin SDE, but rather its ergodicity.

We will therefore discuss tests of BD-HS algorithms that involve probing the hard-sphere interactions but still allow for a comparison with exact results known for the Brownian system, i.e.\ where the Langevin equation can be solved exactly in terms of a time-dependent probability distribution function (PDF). In such a case, an empirical PDF can be generated from sufficiently many runs of the algorithm in question and then be compared to the exact solution.

The methods we are going to discuss in the present paper, are crossbreeds between standard Brownian methods (ignoring the singular potential), and the standard simulation method for the ordinary (non-stochastic) dynamics of hard spheres, called event-driven (ED) simulations. Predecessors date back to Monte Carlo (MC) inspired schemes
\cite{Cichocki90,Heyes93,Moriguchi95,Sillescu94,Foss2000},
culminating in the algorithm by Strating \cite{Strating99}, and algorithms making use of so-called overlap potentials \cite{Tao05}.
Strating's algorithm is but a step away from De~Michele's algorithm \cite{Foffi05}. A slightly different principle has also recently been implemented by Tao \textit{et~al.} \cite{Tao06}, albeit applied to a more complex system of hard rods. We will discuss both these event-driven Brownian dynamics (ED-BD) methods, with an emphasis on De~Michele's algorithm, which we will show to be exact to first order in the integration-step size $\Delta t$.

This paper is organized as follows: the next section discusses the theoretical background, introduces  De~Michele's algorithm and extends it to the case of shear and constant forces. Section \ref{sec:twobody} presents numerical tests for two-particle cases where a comparison with the exact PDFs is possible. In section \ref{sec:manybody} we compare the diffusive many-body behavior with exact theoretical predictions. Section \ref{sec:conclusion} summarizes and concludes.

\section{Hard-Sphere Brownian Dynamics Algorithms}\label{sec:HSBDalgos}

We start from the $N$-particle ($1\le i\le N$) Langevin equation
\begin{equation}\label{eq:langevin}
  m_i\dot{\vec v}_i = \fp_i + \fr_i + \fd_i + \fe_i\,,
\end{equation}
where the subscripts $i$ label the particles and will be dropped where they
are clear from the context.
In Eq.~\eqref{eq:langevin}, $\fp$ incorporates the effect of the HS interaction and $\fe$ is a sufficiently smooth external force. $\fr$ and $\fd$ are the random and dissipative forces resulting in Brownian dynamics, therefore the above equation is a second-order SDE in the particle positions. The random force is characterized by its zero mean and the correlation
\begin{equation}
  \langle\fr(t)\otimes\fr(t')\rangle = 2\kT\bs R\delta(t-t')\,,
\end{equation}
and the fluctuation-dissipation theorem (FDT) then requires the dissipation to be $\fd=-\bs R(\vec v-\vec u)$, where $\vec u$ is a local flow velocity field; $\vec u$ is nonzero in the case of applied shear, $\vec u=\dot{\bs\Gamma}\vec r$, where $\dot{\bs\Gamma}$ is a shear-rate tensor. 
$\bs R$ is in general a complicated matrix depending on the full configuration of the system at any given time, representing hydrodynamic interactions (HI). Here we are concerned with the simpler case where $\bs R=\xi\bs1$ is constant, with a real number $\xi>0$ characterizing the noise.
In this paper we furthermore deal with the limit $m/\xi\to0$, the case of strong dissipation, where Eq.~\eqref{eq:langevin} reduces to
\begin{equation}\label{eq:brownian}
 \xi\vec v =  \fp_i + \fr_i + \xi\vec u + \fe_i\,,
\end{equation}
that is a first-order SDE in the position of the particles. 

In what follows, we will, according to customary procedure \cite{ATBook}, assume that one can fix a small enough time interval $\Delta t$ over which the particle configuration and all smooth forces vary slowly, allowing them to be treated as constant. If this were possible for $\fp$ as well, one could use conventional SDE integrators \cite{KloedenPlatenBook}, a topic in its own right (see e.g.\ Refs.~\cite{Ermak,Xue90,Turq77,Helfand78,Iniesta90,Honeycutt92,Branka98, Skeel02,Wang03,Ricci03}). For hard spheres, the forces are not Lipschitz continuous and standard numerical integrators are ill-defined \cite{KannanBook}.
Integrating Eq.~\eqref{eq:brownian} over the interval $\Delta t$, one obtains
\begin{equation}\label{eq:bddispl}
\Delta\vec r(\Delta t)=\vec g\,\Delta t+\vec\mu(\Delta t)\,,
\end{equation}
where $\vec g=\vec u+\xi^{-1}(\fp+\fe)$ contains the systematic and interaction terms, while $\vec\mu$ is a Wiener process with $\langle\vec\mu(t)\otimes\vec\mu(t')\rangle=2\bs D\min(t,t')$, where $\bs D=\kT{{\bs R}^{-1}}^T=(\kT/\xi)\bs1$ is the matrix of diffusion coefficients. Note that we still allow these bare diffusion coefficients to depend on the particle index, as it is the case in multi-component (polydisperse) systems.

We discuss the case $\vec u=\fe=0$ first. One approach is to simply set $\fp=0$ in generating random displacements $\Delta\vec r$ with the statistics given by $\vec\mu$ \cite{Cichocki90,Heyes93,Moriguchi95,Sillescu94,Foss2000,Strating99}. A second step then is needed to ensure that the random displacements are compatible with the presence of $\fp\neq0$; for hard spheres this is the criterion that no two particles overlap. One can either discard all displacements that violate this condition -- this is a variant of standard Metropolis MC (called simply MC in the following). It has been shown \cite{Cichocki90} that in the limit $\Delta t\to0$ one indeed obtains Brownian dynamics, i.e., a faithful realization of Eq.~\eqref{eq:brownian}. We will discuss this point again further on.

More elaborate overlap removal can be applied: some have put particles back at contact (or around contact on average) along the line of their relative displacement \cite{Heyes93,Moriguchi95,Sillescu94,Foss2000}, although this has unwanted effects on the pair distribution function, hence on observables like pressure. Strating \cite{Strating99} proposed to remove overlaps by performing elastic collisions: from the time step $\Delta t$ and the displacement $\Delta\vec r$, assign a fictive velocity $\vec v=\Delta\vec r/\Delta t$ to the particles, and move them by displacements $\Delta\vec r^{\,'}$ as if they underwent ballistic flight (including eventual elastic collisions) with this velocity $\vec v$ in the time interval $\Delta t$.

Whatever the removal procedure, one has to be aware that the removal of one overlap can create another, called secondary (tertiary, and so forth) overlaps. This is especially likely to happen in dense systems. As Strating pointed out, both the removal of all these `higher-order' overlaps as well as the order in which overlaps are removed are crucial for the algorithm to work. According to Ref.~\cite{Strating99}, one needs to remove them in the order in which they would have occurred in a Newtonian dynamics simulation.

At this point it is convenient to turn around the discussion and start
from an algorithm that treats Eq.~\eqref{eq:brownian} in the opposite
limit: If $\xi$ were zero, one would recover the standard Newtonian dynamics of hard spheres. The best numerical schemes in this case are event-driven (ED) simulations \cite{RapaBook,Allen89}: assuming the collisions are binary in all non-degenerate cases and of infinitesimal duration, one advances the system from one such collision to the next, solving the ballistic free flight in between exactly. The random force in Eq.~\eqref{eq:brownian} prevents a na\"{\i}ve application of this approach, but at least every $\Delta t$ we can reintroduce the randomness: if we interpret the velocities of the ED simulation as the fictive velocities of the above discussion, randomly drawn every $\Delta t$, the ED scheme is simply a device to prevent all unphysical overlaps in the first place, using a set of vectors $\vec v$ as its book-keeping device.

Note that in terms of efficiency, there is little difference to Strating's method, since the main effort in ED simulation goes into the calculation and the sorting of collision times, something that is also needed when one is to remove overlaps in their `correct' order. Note also that this algorithm bears a resemblance to some ED granular matter simulations \cite{Luding94,Luding95,Williams96,Peng98,VanNoije99}, where however a modified kinetic equation instead of Eq.~\eqref{eq:langevin} needs to be solved.

The question then arises how to handle `Brownian collisions' in such a combined ED-BD scheme. We will argue below that the elastic collision rule of Newtonian dynamics is a reasonable choice, inspired by its exactness in one dimension. This is De~Michele's algorithm, first used in \cite{Foffi05}. In the limit of small $\Delta t$ and for vanishing $\vec u$ it is essentially Strating's algorithm, with the overlap problem cured. (For large $\Delta t$, the two algorithms differ in the treatment of one particle `tunneling' across another.) 
We we also discuss another choice, made by Tao \textit{et~al.} \cite{Tao06} in the more complicated context of hard rods: instead of ballistic collisions, where the pre- and post-collision velocities are perfectly correlated, one can decorrelate them by assigning the post-collision velocity randomly (with proper restrictions to avoid overlaps).



To investigate the role of ED collisions in the BD algorithm, we study the two-body BD-HS problem without external forces in more detail. This can be solved analytically \cite{Hanna82,Ackerson82}, by transforming from the particle coordinates $\vec r_i$, $i=1,2$, to the relative, $\vec r=\vec r_1-\vec r_2$, and center-of-motion coordinates, $\vec R=\Xi^{-1}(\xi_1\vec r_1+\xi_2\vec r_2)$, where $\Xi=\xi_1+\xi_2$. The latter then separates from the problem, giving free center-of-motion diffusion. In the relative coordinates, the hard-core interactions can be included as a boundary problem into the corresponding diffusion equation,
\begin{subequations}
\begin{align}
  \partial_t p(\vec r,t)&=D\nabla^2p(\vec r,t)\,,&r>\sigma\,,\\
  \vec r\cdot\vec\nabla p(\vec r,t)&=0\,,&r=\sigma\,,
\end{align}
\end{subequations}
where $p(\vec r,t)$ is the PDF for finding a relative distance $\vec r$ at time $t$, $D=D_1+D_2$ is the relative diffusion coefficient and $\sigma=(\sigma_1+\sigma_2)/2$ with the particle diameters $\sigma_i$.

One could aim to include the exact solution of this case in an HS-BD algorithm, by drawing particle displacements according to this $p(\vec r,t)$ whenever two particles are sufficiently near. However, the procedure would be rather involved, since the analytical solution \cite{Hanna82,Ackerson82} is only available as an infinite sum in the Laplace domain. It would also be only approximate in the presence of more than two particles, hence valid only for small time steps, and again introduce secondary overlap problems.

It is therefore easier to consider the limit of small $\Delta t$;  then, the boundary at $r=\sigma$ can be considered flat, and the analytical PDF is easily obtained by the method of images \cite{PolyaninBook}: denoting by $G_0(\vec r,t_0+\Delta t|\vec r_0,t_0)$ the Green's function of the unbounded diffusion equation (for a particle starting at $\vec r_0$ at time $t_0$), we have
\begin{multline}
  p(\vec r,t_0+\Delta t|\vec r_0,t_0)\\ =
     G_0(\vec r,t_0+\Delta t|\vec r_0,t_0)
    +G_0(\vec r,t_0+\Delta t|\vec r_0^{\,*},t_0)
  \label{mirrorG}
\end{multline}
for $r>\sigma$, and zero otherwise.
Here, $\vec r_0^{\,*}$ is the mirror image of the initial coordinate $\vec r_0$ with respect to the boundary. An algorithm which implements transitions of a particle from $\vec r_0$ to $\vec r$ in the time interval $\Delta t$ by performing a ballistic free flight with a velocity $\vec v$ chosen from some probability distribution $f(\vec v)$ that satisfies $f(\vec v = \Delta\vec r/\Delta t)d\vec v =  G_0(\Delta\vec r,t_0+\Delta t|0,t_0)d\vec r$, will satisfy the correct diffusion equation for such no-flux boundary conditions. This holds as long as the correct solution can be obtained by the image method, and if $G_0$ describes a homogeneous process, i.e., depends on position only through the difference $\Delta\vec r=\vec r-\vec r_0$, and not on $\vec r$ and $\vec r_0$ individually.

The fictive-velocity distribution corresponding to free Brownian diffusion,
$G_0(\Delta\vec r,\Delta t)=(4\pi D\Delta t)^{-d/2}
\exp[-(\Delta\vec r)^2/(4D\Delta t)]$,
is of course just the Maxwell-Boltzmann distribution,
$f(v)=(\beta m/2\pi)^{-d/2}\exp[-\beta(m/2)v^2]$. This connection is essentially the way in which the FDT enters the ED-BD algorithms. The parameter $\beta m_i$ sets the bare diffusion coefficient for each particle $i$, given as
\begin{equation}\label{eq:diffm}
D_i=\frac{\Delta t}2\frac1{\beta m_i}\,.\end{equation}
In particular, polydisperse systems, where different HS species have different $D_i$, can be simulated by incorporating elastic collisions according to the (fictive) masses $m_i$.

De~Michele's algorithm \cite{Foffi05} for hard spheres then reads:
(i) every $t_n=n\Delta t$ ($n$ integer) extract velocities $\vec v_i$
according to a Maxwellian distribution with fictive masses obeying
Eq.~\eqref{eq:diffm}; 
(ii) evolve the system between $t_n$ and
$t_n+\Delta t$ according to the laws of ballistic motion (performing
ED molecular dynamics). 

As outlined above, one can anticipate that this algorithm converges to the correct solution in the limit $\Delta t\to0$, where the problem reduces to the one-dimensional two-body case, which is treated exactly. 
A first estimate for its regime of applicability is $\Delta t\ll\sigma^2/(4D)$ -- in order to treat particle boundaries as flat walls -- and $\Delta t\ll d^2_{\text{avg}}/(4D)$, where $d_{\text{avg}}\sim\rho^{-1/3}$ is a typical interparticle separation -- so that `almost all' time steps deal with binary interactions only.

The image-method solution can be extended in a straightforward manner to include a linear shear field, $\vec u=\dot{\bs\Gamma}\vec r$, and external forces that change slowly in space (such that they can be approximated as constant over the typical $\Delta r$ resulting in a time step).
Let us discuss the case of linear shear flow in more detail. In this case one needs to deal with \emph{two} effects on the Brownian motion: there will be a systematic drift $\vec u$, but also the noise term for $\Delta r$ will be modified because $\vec u$ depends on $\vec r$. If the shear flow acts along the $x$-direction, $\vec u=\dot\gamma y\vec e_{\vec x}$, we have \cite{Foister80}: $\langle\Delta y\rangle=\langle\Delta z\rangle=0$, $\langle\Delta x\rangle=y\dot\gamma\Delta t$, and $\langle\Delta y^2\rangle=\langle\Delta z^2\rangle=2D\Delta t$, but $\langle\Delta x^2\rangle=\langle\Delta x\rangle^2+2D\Delta t(1-(1/3)(\dot\gamma\Delta t)^2)$. The random displacements appearing in Eq.~\eqref{eq:bddispl} in the presence of linear shear are also cross-correlated, $\langle\Delta x\Delta y\rangle=(D\Delta t)(\dot\gamma\Delta t)$.

This leads to the following extension of De~Michele's algorithm:
(i) extract random velocities 
$\vec v = \left( \Delta\vec r - \langle\Delta\vec r\rangle\right)/\Delta t$ 
from a multivariate Gaussian distribution according to the above averages, 
(ii) add to the fictive velocities the systematic drift 
$\langle\Delta\vec r\rangle/\Delta t$ 
induced by the shear flow and/or gravity.
In previous applications of HS-BD algorithms to sheared systems, only the
systematic drift has been taken into account (cf.\ Ref.~\cite{Strating99}).
The effect of neglecting these high-shear-rate corrections to the
Brownian displacements has not been studied so far and remains to be
clarified.
For constant external forces (such as gravity), only a systematic drift needs to be taken into account.

\section{Two-Particle Tests}\label{sec:twobody}

As in two (and higher) dimensions, no exact solution in terms of a mirror image exists for the two-body HS problem, De~Michele's  algorithm introduces an approximation which is worthwhile testing. The exact distribution function in two dimensions is, after a Laplace transform from time $t$ to frequency
$s=D^2q^2$ \cite{Ackerson82}
\begin{multline}\label{eq:2dimpdf}
  p(r,\phi,s|r_0,\phi_0)=\frac1D\sum_{m=-\infty}^\infty
  e^{im(\phi-\phi_0)}/(2\pi) K_m(qr_>)\\ \times
  \left[I_m(qr_<)-K_m(qr_<)\frac{I_m'(q\sigma)}{K_m'(q\sigma)}\right]\,,
\end{multline}
where we have transformed to polar coordinates $(r,\phi)$. Here, $r_>$ is the greater of $r,r_0$, and $r_<$ the lesser. $I_m(z)$ and $K_m(z)$ are the modified Bessel functions of the first and second kind of order $m$. Note that here and in the following, $D$ denotes the \emph{relative} diffusion coefficient between the two spheres.
On the other hand, the PDF of De~Michele's algorithm, Eq.~\eqref{mirrorG}, is in two dimensions Laplace transformed using $\LT[1/(4\pi Dt)\exp[-\alpha^2/(4Dt)]=1/(2\pi)K_0(\alpha q)$. The latter may be expressed using the addition theorem for the modified Bessel functions \cite{CourantHilbertBook} as
\begin{multline}
  p_{\text{ED}}(r,\phi,s|r_0,\phi_0)=\frac1D\sum_{m=-\infty}^\infty
  e^{im(\phi-\phi_0)}K_m(qr_>)\\ \times
  \left[I_m(qr_<)+K_m(qr_>^*)\frac{I_m(qr_<^*)}{K_m(qr_>)}
  e^{im(\phi_0-\phi_0^*)}\right]\,,
\end{multline}
where $(r_0^*,\phi_0^*)$ are the angular coordinates of the mirror image. Since they depend in a complicated fashion on $(r,\phi)$, it is hard to assess the error made in this approximation analytically.
At least in the limit $q\to\infty$, one recognizes from
$I_m(qr_<^*)/K_m(qr_>)\sim-I_m'(qr_<^*)/K_m'(qr_>)$, that
$p_{\text{ED}}(r,\phi,s)$ approaches the true $p(r,\phi,s)$ in the limit
$r_0\to\sigma$.

We can however proceed by looking at a particular average that can be calculated exactly. Let us consider the average displacement magnitude $|\langle\Delta\vec r\rangle_{d\text{D}}| =(\sum_i\langle\Delta x_i\rangle_{d\text{D}}^2)^{1/2}$ where $x_i$ label the Cartesian components and $\langle\Delta x_i\rangle_{d\text{D}}=\int_{r>\sigma}r^{d-1}drd\Omega_d\Delta x_ip(\vec r,s|\vec r_0)$ is the $d$-dimensional average. With the above form of the two-dimensional exact PDF, Eq.~\eqref{eq:2dimpdf}, this average evaluates to a relatively simple expression (see Appendix \ref{appendixLT}). In three dimensions, the calculation proceeds along the same lines, essentially replacing the Bessel functions $I_m(z)$ and $K_m(z)$ by their spherical counterparts, $i_m(z)$ and $k_m(z)$, and involving spherical harmonics in place of the angular exponentials. We get
\begin{align}
  \label{eq1dimavg}
  \langle r\rangle_{\text{1D}}
  &=\frac1{q^3}e^{-q(r_0-\sigma)}\,,\\
  \label{eq2dimavg}
  \left|\langle\vec r\rangle_{\text{2D}}\right|
  &=\left|\LT^{-1}\left[\frac1{q^3}\frac{K_1(qr_0)}{K_1'(q\sigma)}\right]
  \right|\,,\\
  \label{eq3dimavg}
  \left|\langle\vec r\rangle_{\text{3D}}\right|
  &=\left|\LT^{-1}\left[\frac1{q^3}\frac{k_1(qr_0)}{k_1'(q\sigma)}\right]
  \right|\,,
\end{align}
always using $r_0>\sigma$. The one-dimensional result can also be obtained by recurring to the corresponding Laplace-transformed PDF \cite{Ackerson82}, or obtained in the time domain by direct integration (making use of the image-method solution).

Note that the two- and three-dimensional expressions for large $q$ have Eq.~\eqref{eq1dimavg} as their leading asymptote. The same holds for the PDF implemented by De~Michele's algorithm, again demonstrating that this method converges to the correct two-body solution for small enough $\Delta t$. The rest of the paper will basically deal with the question of how small $\Delta t$ needs to be in order for the method to produce meaningful results.

\begin{figure}
\includegraphics[width=.7\figurewidth]{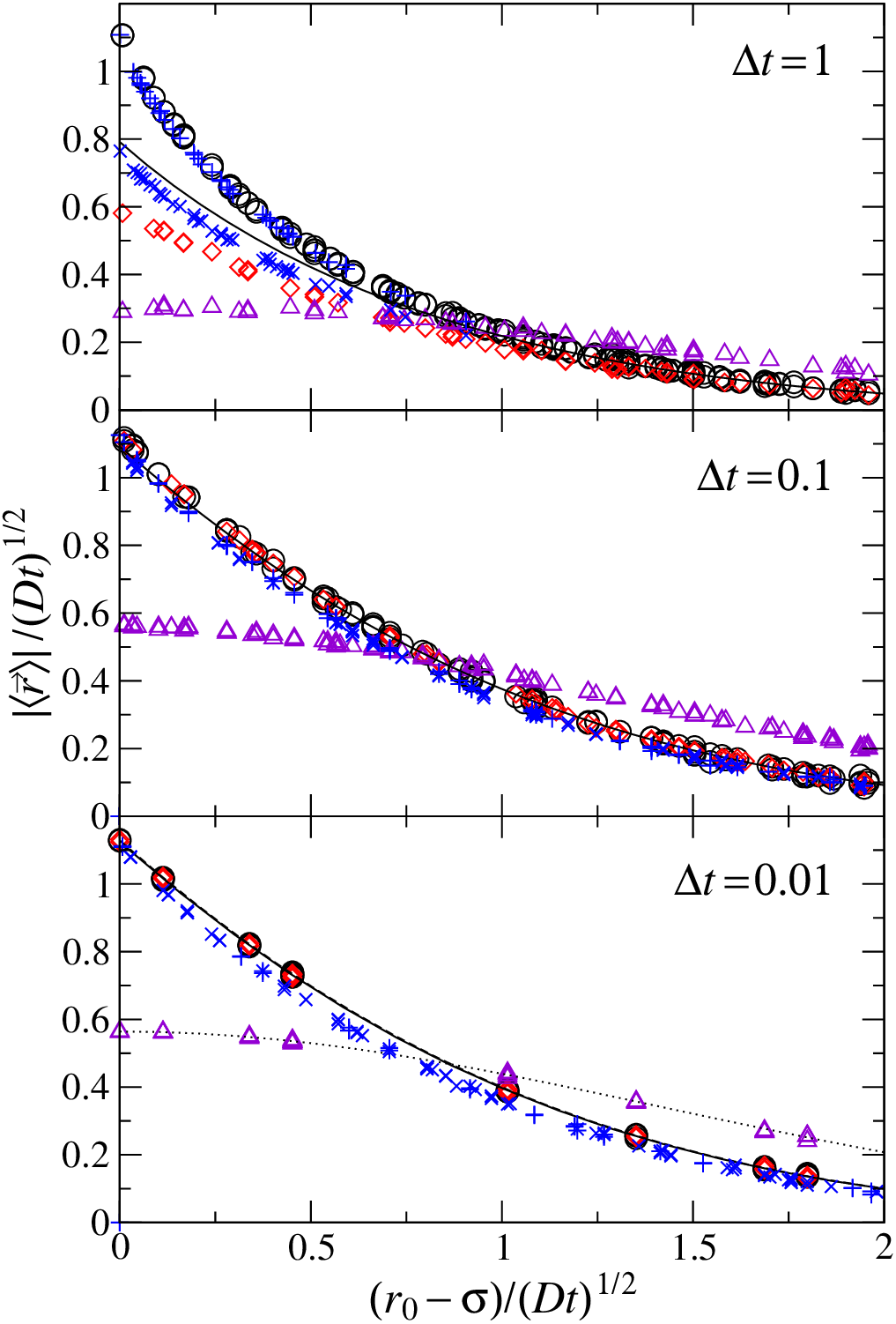}
\caption{\label{fig:2davg}
Average displacement size, $|\langle\Delta \vec r\rangle_{\text{2D}}|$, for the two-dimensional Brownian motion of two equal hard spheres ($\sigma=1$) in their relative-coordinate frame; $r_0$ is the initial center-to-center distance. The relative diffusion coefficient is $D=\Delta t/2$; the step size $\Delta t$ is kept fixed within each panel. All values are scaled by $1/\sqrt{D\Delta t}$. Symbols are one-timestep simulation results, averaged over $50000$ runs for each point: De~Michele's algorithm (circles), the Tao-BD algorithm (plus symbols), a modified Tao-BD algorithm (crosses), Strating's algorithm (diamonds), and a MC algorithm (triangles). For $\Delta t=0.01$, diamonds and circles, as well as crosses and plus symbols overlap. Solid lines are the analytical solutions. The dotted line corresponds to $\exp[-(r_0-\sigma)^2/(4Dt)] /\sqrt{\pi Dt}$.
}
\end{figure}

To assess the convergence of the algorithm, we have numerically inverted the Laplace transform of $\langle\vec r\rangle$, Eqs.~\eqref{eq2dimavg} and \eqref{eq3dimavg}, for finite $\Delta t$ using a publicly available Gaver-Wynn~Rho algorithm \cite{LTinversion}. In the following, we set units such that $D=1$ and $\sigma=1$. The exact solutions are compared with averages obtained from simulation runs in two dimensions in Fig.~\ref{fig:2davg}. Each data point in the figure corresponds to an average over $50000$ runs starting from a given $(x_0,y_0)$, each performing one step of size $\Delta t$ as indicated in the different panels of the figure.

De~Michele's algorithm (circles) and Strating's algorithm (diamonds) give almost identical results for small enough $\Delta t$; only at unrealistically large $\Delta t$ does one observe a difference due to `missed' collisions in Strating's algorithm. At this point, however, both algorithms already deviate significantly from the exact solution (shown as a solid line). The convergence to the right solution is good enough to give reasonable results already for $\Delta t=0.1$, about one order of magnitude bigger than what has been used in previous studies \cite{Strating99,Foffi05,Tao06}. For even smaller $\Delta t$, also the exact solution becomes indistinguishable from the one-dimensional
$\langle\Delta r\rangle_{\text{1D}}$, the time-domain version of Eq.~\eqref{eq1dimavg}:
\begin{equation}\label{eq1dimavgt}
  \langle\Delta r\rangle_{\text{1D}}=\Theta(\rho_0)
  \frac{2\sqrt{D\Delta t}}{\sqrt\pi}
    e^{-\rho_0^2/(4D\Delta t)}-\rho_0\erfc\left(\frac{\rho_0}
    {\sqrt{4D\Delta t}}\right)
\end{equation}
(where $\rho_0=r_0-\sigma$).

Let us point out that this convergence is indeed closely related to the elastic-collision rule. In fact, from a Metropolis-MC scheme with Gaussian displacements \cite{Cichocki90}, one gets the results shown as triangles in the figure, approaching for small $\Delta t$ its one-dimensional limit
\begin{equation}\label{eq1dimavgt_mc}
  \langle\Delta r\rangle_{\text{1D,MC}}=\Theta(\rho_0)
    \frac{\sqrt{D\Delta t}}{\sqrt\pi}
         e^{-\rho_0^2/(4D\Delta t)}\,.
\end{equation}
Here we have dropped a term due to the possible `tunneling' of the particle across the excluded-volume region, which is exponentially small for $\Delta t\ll\sigma^2/(4D)$. The two expressions, Eqs.~\eqref{eq1dimavgt} and \eqref{eq1dimavgt_mc}, differ in their leading-order terms by a factor of two. This highlights that the approach of the MC dynamics to true Brownian dynamics even as $\Delta t\to0$ is less straightforward than one might suppose: the MC method neglects a term of $\mathcal{O}(\sqrt{\Delta t})$ that is part of the finite-time-step solution of the Smoluchowski equation, whereas De~Michele's algorithm has a leading error term of $\mathcal{O}(\Delta t)$.
A similar argument has been brought forward by Heyes and Bra\'nka \cite{Heyes98} for the case of smooth forces. There, however, one needed to look at the mean-squared displacement, $\langle\Delta r^2\rangle$ to find differences at finite $\Delta t$, whereas here the disagreement sets in one level earlier.
It also explains why the use of Monte Carlo to simulate BD needed an elaborate
density-dependent extrapolation procedure to $\Delta t=0$, using several
simulation runs at different $\Delta t>0$ \cite{Cichocki90}.

The Tao \emph{et~al.} \cite{Tao06} algorithm is more difficult to assess. It replaces the deterministic reflection of the trajectory by a stochastic collision law and can for hard spheres be formulated as follows: assigning a fictive velocity to the particle from the random displacement $\Delta r$, one calculates the time $0\le t_c\le\Delta t$ for which $x(t_c)=\sigma$ is reached, and then assigns a final position from new random displacements in accord with the remaining time $\Delta t-t_c$. The procedure yields in one dimension
\begin{multline}
  p_{\text{Tao,1D}}(r,t_0+\Delta t|r_0,t_0)
  =G_0(r,t_0+\Delta t|r_0,t_0)\\
  +\int_{-\infty}^{(\sigma-r_0)/\Delta t}
  \frac{\Delta t e^{-v^2/(4D\Delta t)}}{\sqrt{4\pi D\Delta t}}
  \frac{2 e^{-(r-\sigma)^2/(4D\Delta t(1-\eta))}}
  {\sqrt{4\pi D\Delta t(1-\eta)}}
  \,dv\,,
\end{multline}
where $\eta=(\sigma-r_0)/(v\Delta t)$.
The first term in the integral is the probability of assigning a fictive velocity $v$ to the particle such that a collision takes place at $t_c=(\sigma-r_0)/v$, while the second term is the probability to arrive at the final point $r$ by a random displacement starting from $r(t_c)=\sigma$ and constrained to be directed away from the wall. Indeed, as $r_0\to\sigma$, we have $\eta\to0$, and the second term in the above integral gives the correct image-point contribution, while the factor $2$ cancels with the $v$-integration over the first term. Apart from this asymptotic case, it is however difficult to evaluate $p_{\text{Tao}}$ analytically. Numerical evaluation shows it to be less accurate if $(r_0-\sigma)/\Delta t$ is not small but of intermediate magnitude.
There is an ambiguity in the translation of the Tao-BD recipe
described in Ref.~\cite{Tao06} when the colliding objects have curved
surfaces, regarding how the random post-collision velocities should be
generated. Tao \textit{et~al.}\ state that they should be such that the
particles separate initially. If one implements this algorithm for hard spheres,
the results for $\langle\Delta\vec r\rangle$ are almost identical to
De~Michele's ones for large $\Delta t$, as the plus symbols in
Fig.~\ref{fig:2davg} show. For small $\Delta t$, a small but discernible
difference always remains, owing to the randomness in the velocity reflection. One can introduce
a slight modification to the Tao-BD algorithm, in which the post-collision
velocities are only required to be such that the two colliding particles do
not overlap at the end of the time step. If one does so, one can improve
on the result for $\langle\Delta\vec r\rangle$, as the crosses in
Fig.~\ref{fig:2davg} show: for small $\Delta t$, this algorithm behaves
just like the original Tao-BD one, as expected, and for larger $\Delta t$,
its results remain closer to the true solution. We did, however, not
test this modification for the many-particle case: relaxing the criterion
for the post-collision velocities can in principle lead to the same
secondary-overlap problems the Strating algorithm suffers from.


Having discussed the one-step behavior of the algorithm, let us now look at the
results after many steps $M$, i.e., at a time $T=M\Delta t$ large compared to
$\Delta t$, $M\gg1$. Here we perform a numerical comparison of the one-particle
PDF in the two-dimensional case in the presence of fixed other spheres
(corresponding to the relative part of the PDF in the two-body problem).
We restrict the discussion to De~Michele's algorithm now.
As initial condition, let us fix a relative distance $r_0=1.1\sigma$.
Since we are interested in the PDF as a function of $t$, and the exact solution
of this two-body case in Refs.~\cite{Hanna82,Ackerson82} is given as a infinite
series in Laplace-transformed frequency $s$, we found it easier to solve for
the PDF numerically. 
Using Crank's method \cite{CrankBook} on a 2-dimensional $K\times K$ periodic
square mesh ($K=800$) of length $L=8$, conditions of no flux are imposed on the
circle; with such conditions Crank's method conserves the probability.
Indicating with $\delta x=L/K$, Crank's method is stable for time steps
$D\,\delta t\ll \delta x^2$; the integration time step was chosen as
$\delta t=10^{-2}\delta x^2/D$, checking
that smaller time steps do not improve the accuracy of the
numerical solution. The simulation data was obtained from averages over
$200000$ runs, then using data binning with $dx=dy=0.32$ for both the
simulation data and the numerically solved PDF.

\begin{figure}
\includegraphics[width=\figurewidth]{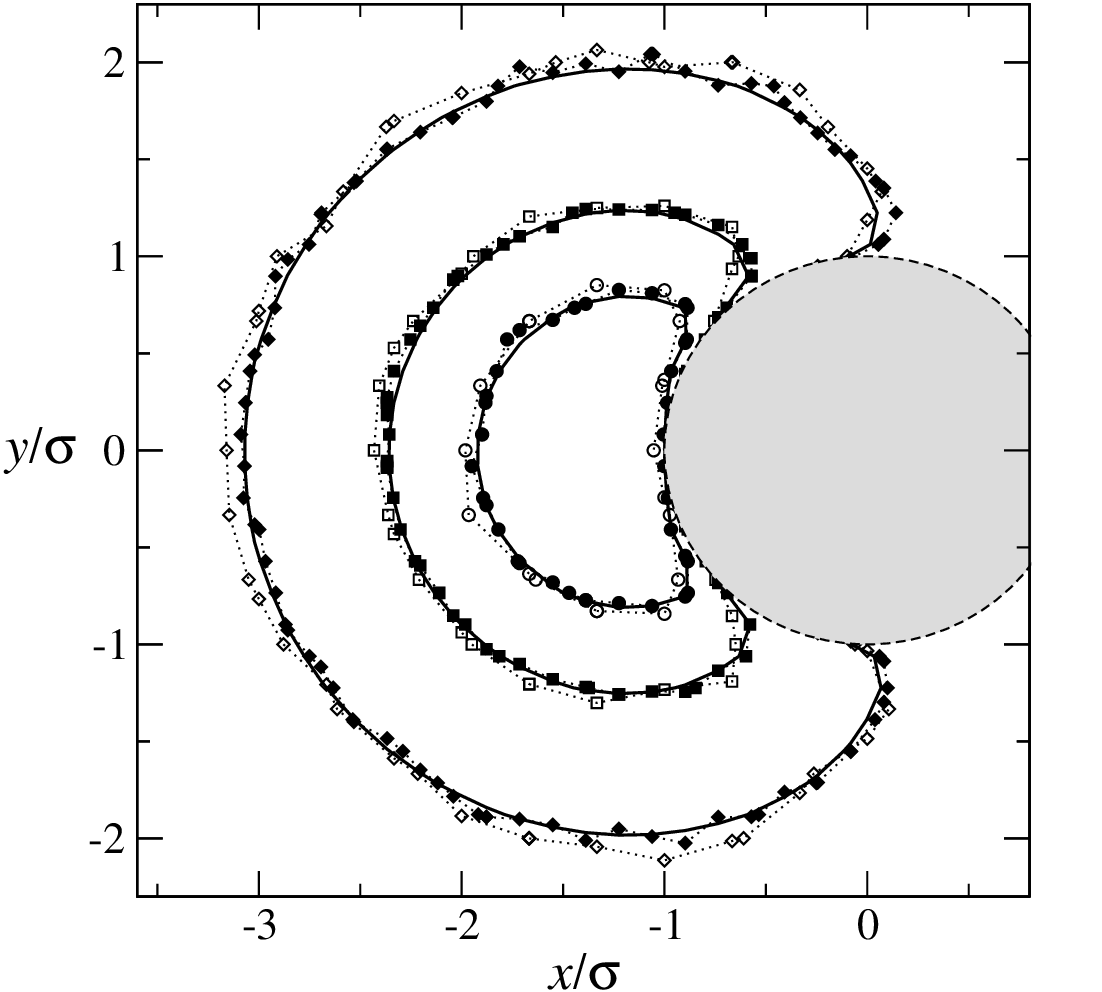}
\caption{\label{fig:lr1contour}
Contour plot comparing the probability distribution (PDF) for the relative diffusion of two hard spheres (diameter $\sigma=1$, relative diffusion coefficient $D=1$), at time $T=1/4$.
Circles (squares, diamonds) connected with dotted lines are the positions where the binned simulation-data PDF has reached its half-maximum ($1/5$, $1/50$); filled symbols correspond to $\Delta t=0.01$ ($5000$ steps), open symbols to $\Delta t=0.1$ ($50$ steps). Solid lines are the corresponding exact-PDF results. A bin size of $dx=dy=0.32$ was used. The grey circle indicates the position of the second particle (no-flux boundary condition along the dashed line).
}
\end{figure}

A visual inspection already yields some insight on the quality of the agreement. To this end, we plot equal-probability contour lines corresponding to the half-, 1/5-, and 1/50-maximum value of the PDF. The comparison of the exact result with De~Michele's algorithm is shown in Fig.~\ref{fig:lr1contour}. There, solid lines indicate the exact results, while the different symbols represent the results of De~Michele's algorithm with different time steps, $\Delta t=0.01$ (corresponding to $M=5000$ steps) and $\Delta t=0.1$ ($M=50$). The fluctuations visible in the figure are all well within the fluctuations expected for the number of runs used to obtain the average.

While the above discussion refers to a two-dimensional test, the same
features hold in three dimensions. There, a simple expression for the
exact two-particle PDF can be obtained, if one starts from an angular-averaged
initial distribution $P(\vec r,0|\vec r_0)=\delta(r-r_0)/(4\pi r_0^2)$, viz.\
\cite{SchultenEbook}:
\begin{multline} \label{eq:prob3dexact}
  4\pi r r_0\, p(r,t|r_0) = 
  \\ \frac{1}{\sqrt{4\pi Dt}}
  \left(\exp\left[\frac{-(r-r_0)^2}{4Dt}\right]
  + \exp\left[\frac{-(r-r_0^*)^2}{4Dt}\right] \right)\\
 -\frac{1}{\sigma}
  \exp\left[ \frac{Dt}{\sigma^2} + \frac{r-r_0^*}{\sigma}\right]
 \erfc\left[ \sqrt{\frac{Dt}{\sigma^2}} + 
  \frac{r-r_0^*}{\sqrt{4Dt}}\right]
\end{multline}
(where $r_0^*=2\sigma-r_0$).

\begin{figure}
\includegraphics[width=\figurewidth]{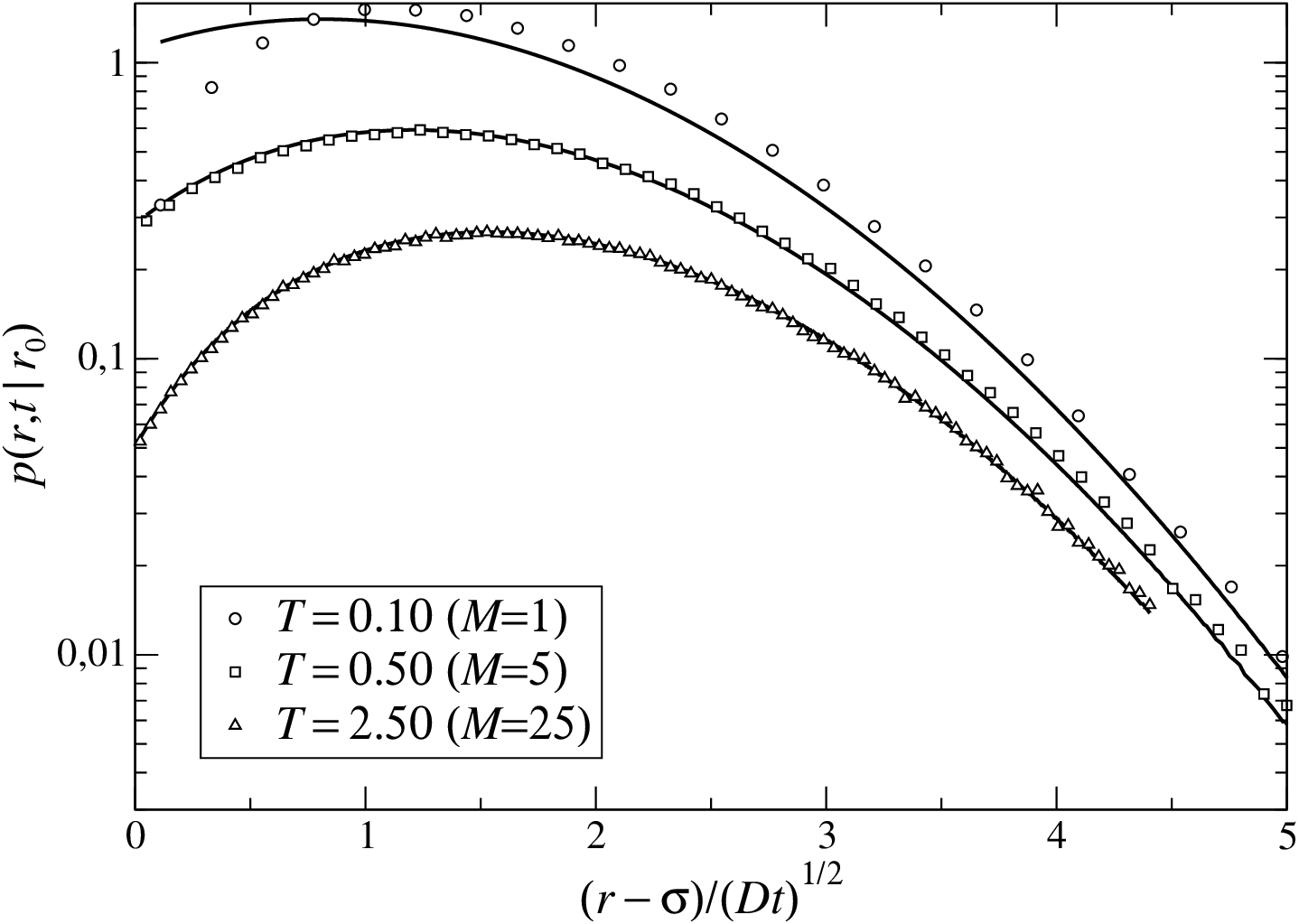}
\caption{\label{fig:pdftest3d}
Two-particle reduced PDF in three dimensions (with angular-averaged initial conditions), $p(r,t|r_0)$, Eq.~\eqref{eq:prob3dexact}, at times $T=0.1$, $0.5$, and $2.5$, as a function of the scaled initial separation $(r_0-\sigma)/\sqrt{Dt}$, where $r_0=1.01\sigma$ (solid lines). Results from the De~Michele's algorithm with $\Delta t=0.1$ are shown as symbols; they correspond to $M=1$, $5$, and $25$ simulation time steps.
}
\end{figure}

The reduced PDF, Eq.~\eqref{eq:prob3dexact}, is compared in Fig.~\ref{fig:pdftest3d} with the results from De~Michele's algorithm. We show a cut using $r_0=1.01\sigma$ and a fixed time step $\Delta t=0.1$ in the algorithm. The final time $T$ is varied through the number of simulation time steps $M$. Note that this $\Delta t$, as judged from Fig.~\ref{fig:2davg}, is already quite large, so that for only one simulation time step, a discrepancy is clearly visible. This however vanishes quite quickly; already for $M\approx5$ the deviations are minute, and for $M\approx25$, they are no longer identified. Generally, this indicates that the transition from two to three dimension in the analysis of the algorithm poses no surprises. Furthermore, even if in the one-step comparison some differences remain for moderately large $\Delta t$, the results obtained after $M\gg1$ steps can still be correct.

\section{Many particles case}\label{sec:manybody}

We now turn to a brief check of the many-body behavior of the algorithm.
Here, analytical results are available for the long-time diffusion coefficient $D_L$ in the low-density limit \cite{Hanna82}. One gets $D_L=D(1-2\phi)$, where $\phi=(\pi/6)n\sigma^3$ is the packing fraction, and $n$ the number density of the hard-sphere system. Since this result is derived from the two-body PDF discussed above, it poses a genuine test of BD including hard-core exclusion.

\begin{figure}
\includegraphics[width=\figurewidth]{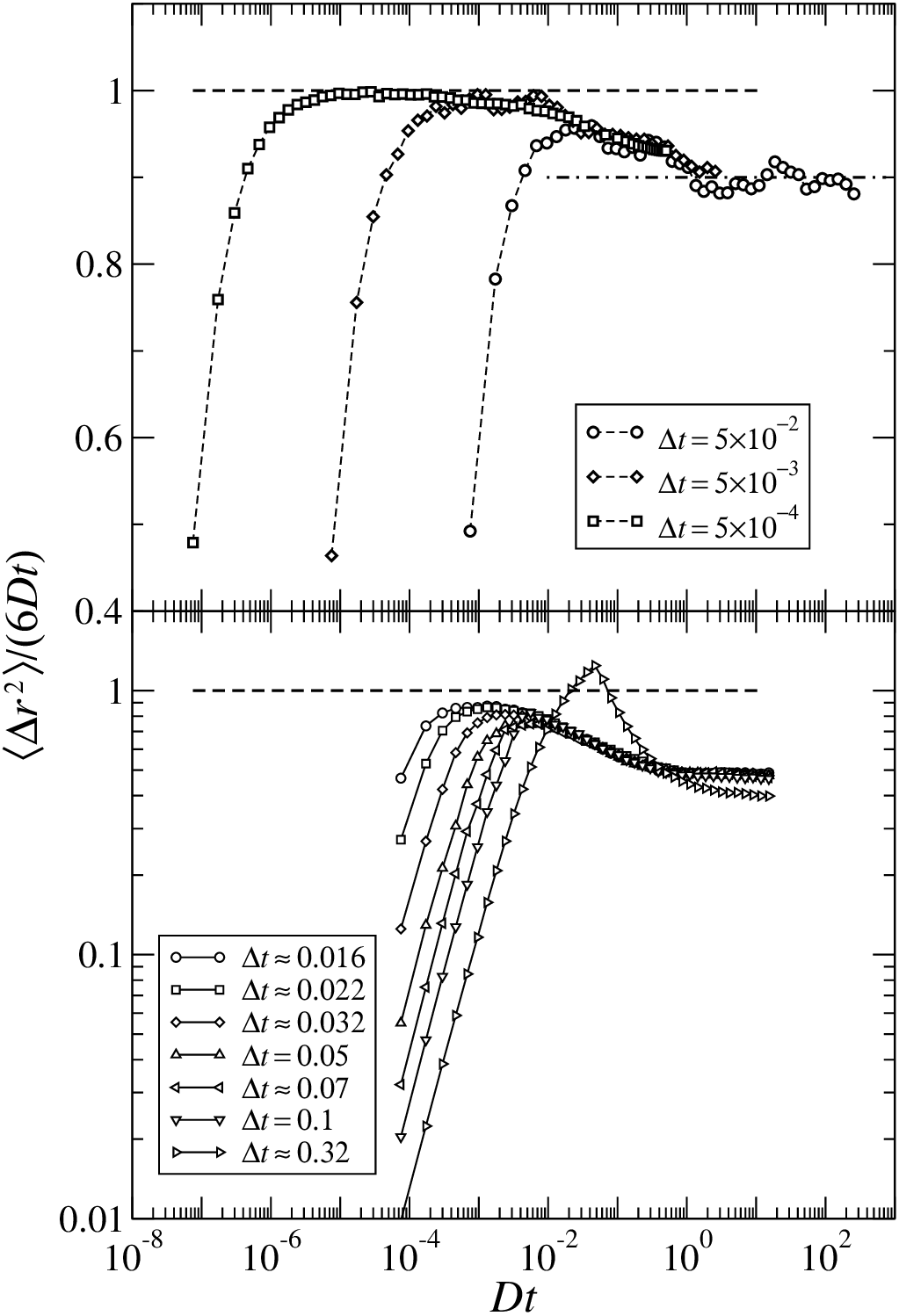}
\caption{\label{Fig:MSD}
Mean-squared displacement $\langle\Delta r^2\rangle$ of hard-sphere systems at packing fractions $\phi=0.05$ (upper panel) and $\phi=0.30$ (lower panel), normalized by the free diffusion limit, $6Dt$.
Symbols are simulations with De~Michele's algorithm using $1000$ particles with various step sizes $\Delta t$ as indicated in the figure.
The dashed horizontal line indicates the short-time asymptote for true Brownian dynamics, $d(t)=\langle\Delta r^2\rangle/(6Dt)=1$. The dot-dashed line corresponds to the long-time asymptote as evaluated in first order in the packing fraction, $d(t)=1-2\phi$.
}
\end{figure}

For the test of De~Michele's algorithm, we used $N=1000$ particles at two different volume
fractions $\phi=0.05$ and $\phi=0.30$.
For each volume fraction several simulations at different time steps $\Delta t$ 
were performed. Results for the mean-squared displacement
(MSD) are shown as symbols in Fig.~\ref{Fig:MSD}, normalized to the free
diffusion asymptote, $d(t)=\langle\Delta r^2\rangle/(6Dt)$. 
The long-time asymptote, $d(t)=1-2\phi$ is reached for $Dt\gtrsim1$, and
this number does not depend on the time step $\Delta t$. The effect of
the finite step size is clearly seen at short times, $\Delta t\lesssim0.05$,
where the ballistic sub-intervals in the algorithm lead to
a MSD quadratic in time, i.e., $d(t)\sim t$.
In agreement with the discussion above (cf.\ Fig.~\ref{fig:pdftest3d}),
the algorithm needs about ${\mathcal O}(10)$ steps in order to reproduce
correctly the Brownian dynamics.
One hence needs to reduce the time
step to $\Delta t={\mathcal O}(10^{-3})$ in order to recover a window in
which the proper Brownian short-time dynamics is visible.
Otherwise, the artificial ballistic
$d(t)$ crosses over directly to the correct long-time behavior, at least
for realistic (not too big) choices of $\Delta t$. It is reassuring that the improper treatment of the short-time dynamics does not influence the convergence to the correct long-time dynamics, and does not even induce an effective time scale (or an effective free diffusion coefficient). In other words, if one is not interested in very early times, a reasonably large $\Delta t$ can be chosen to obtain the dynamics at $t\gg\Delta t$. This is a clear advantage over other schemes, in which an elaborate $\Delta t\to0$ extrapolation needed to be applied. For example, in the MC algorithm we find reasonable agreement agreement for $d(t)$ when Gaussian displacements of variance $\delta r^2={\mathcal O}(0.01\sigma)$ per Cartesian coordinate are chosen. Since in each MC step, a trial move for a single particle is made, this displacement size is connected to a time step by $\delta r^2=2ND\Delta t$ \cite{Cichocki90}.
For our system, $N=1000$, corresponding hence to a time step $\Delta t\ll10^{-7}$ in our units.

As the lower panel of Fig.~\ref{Fig:MSD} shows, the same conclusions can be
drawn from a less dilute case, $\phi=0.3$. Here, the first-order result
quoted for $D_L$ is no longer accurate. Still the $D_L$ read off from the
results of De~Michele's algorithm is stable for $\Delta t\ll0.1$. At $\Delta t=0.1$, a small
deviation remains visible, and only at about
$\Delta t\approx0.3$ (cf.\ the largest step size shown in the figure),
the data show a major deviation also at long times. This data still resembles
the value expected from the dilute case, $d(t)=1-2\phi$, which neglects
three-body terms. The large step size thus violates the condition that
the intervals $\Delta t$ must deal with two-body `events' only, demonstrating
that the restriction $\Delta t\ll d_{\text{avg}}^2/(4D)$ for the maximum
tolerable stepsize indeed becomes the crucial one at moderate and high
densities.

\section{Conclusions}\label{sec:conclusion}

We have discussed schemes to integrate the Brownian motion of a many-body
hard-sphere system, where the singular potential prevents the use of standard
Brownian Dynamics techniques dealing with smooth forces.
Emphasis was placed on testing their correctness
for small but finite integration timesteps $\Delta t$.
We have employed a set of tests based on the exactly known two-particle
probability distribution functions for hard spheres \cite{Hanna82,Ackerson82}.
These tests are sensitive to both the proper implementation of free
diffusion, i.e.\ the random force in the Langevin equation,
Eq.~\eqref{eq:brownian}, and to the treatment of hard-core `collisions',
i.e.\ the singular hard-sphere force $\fp$ in this equation. Hence they
test both crucial ingredients to the HS-BD problem.

It was shown that
De~Michele's ED-BD algorithm indeed converges to the correct solution of Eq.~\eqref{eq:brownian}. The
algorithm works with a finite time step and performs a Newtonian
dynamics simulation within each interval $\Delta t$, where the masses
of the particles play the role of the inverse diffusion coefficients. Every
$\Delta t$, random uncorrelated fictive velocities are assigned to each
particle, used as a book-keeping device to implement MC-like random moves
without overlaps. This realizes the overdamped limit to the Langevin equation.
Unlike a number of earlier schemes (cf.\ Ref.~\cite{Strating99} and citations
therein), this ED-BD method avoids unphysical hard-sphere overlaps in any
case. Methods based on soft-sphere approximations to the hard-sphere
potential need to use very small integration time steps, forced by the
increasing steepness of the potential and the condition that potential
forces vary little during any single time step. Monte Carlo methods,
implementing hard-core exclusion but not flux reflection, reproduce
Brownian dynamics strictly only for $\Delta t\to0$. In contrast, by
implementing the no-flux boundary conditions at least approximately, De~Michele's 
algorithm works reliably with significantly larger time steps. Their magnitude is 
not bound by the steepness of the potential, but only by the size of the
hard spheres and their typical distance. The dynamical features of
Eq.~\eqref{eq:brownian} are correctly reproduced after a small number of
such time steps, and the method is stable in the sense that no drift or
effective rescaling is introduced to the long-time behavior when increasing
the step size within the limits named above. This feature of De~Michele's algorithm 
has been overlooked so far and is an improvement over methods requiring
careful extrapolation to $\Delta t\to0$.

As a simple extension of well-tested event-driven methods for hard spheres,
De~Michele's algorithm shares their effectiveness (but also the problem of being
difficult to parallelize). The same holds for the method recently developed
by Tao \textit{et~al.}\ \cite{Tao06}, although the latter is somewhat more
expensive in terms of computing time: there, considerably more random numbers
need to be drawn (at least one more per collision and dimension), while 
De~Michele's algorithm uses simpler velocity-reflection laws. We found no
considerable difference concerning the results between the two methods.

De~Michele's algorithm can easily be extended from the equilibrium case tested
here. We have discussed in particular how to include linear shear flows (of in
principle arbitrary magnitude). At large shear rate, one needs to take
into account two modifications, namely a deterministic drift and a distortion
of the fictive-velocity distribution. The latter has so far been ignored in
the discussion of BD-HS algorithms.
Further extensions such as the inclusion of finite inertial terms, or of
finite stepwise interactions, will be discussed in
subsequent publications.


\begin{acknowledgments}

We acknowledge support from MIUR-Firb and Cofin. We thank M.~Fuchs for
useful discussion and F.~Sciortino for his ongoing support. 
Th.~V. thanks for the hospitality at the Dipartimento di Fisica of Universit\`a di Roma ``La Sapienza'' where this research has been started and for funding from the Emmy Noether programme of the Deutsche Forschungsgemeinschaft, DFG Vo~1270/1-1.

\end{acknowledgments}


\begin{appendix}

\section{Average relative displacements}\label{appendixLT}

We calculate the average relative displacement of two Brownian hard spheres of
diameter $\sigma$; fixing the relative diffusion coefficient to be $D$. This quantity is most conveniently calculated employing Laplace-transformed time (frequency $s=Dq^2$) and partial integration of the diffusion equation, viz.\ in 1D
\begin{multline}
  \langle r(s)\rangle_{\text{1D}} = (D/q^2)\int_\sigma^\infty r\partial_r
  p(r,s|r_0)\,dx\\
  =(D/q^2) p(\sigma,s|r_0) = (1/q^3)e^{-q(r_0-\sigma)}
\end{multline}
where $p(x,s|x_0)$ is the two-particle PDF in the Laplace domain (see Ref.~\cite{Ackerson82}) and boundary terms vanish due to the no-flux boundary condition.

A similar result holds in 2D: Note that for the Cartesian coordinate $x$,
\begin{multline}
  (q^2/D)\langle x(s)\rangle_{\text{2D}}
  =\int_\sigma^\infty dr\int_0^{2\pi}d\phi\left[r\partial_r(r\partial_r p)
  +\partial_\phi^2p\right]\cos\phi\\
  =\int_0^{2\pi}d\phi\left[\sigma p(\sigma,\phi,s|r_0,\phi_0)\cos\phi\right]\,.
\end{multline}
Replacing $\cos\phi$ by $\sin\phi$ yields $\langle y(s)\rangle$.
Inserting Eq.~\eqref{eq:2dimpdf},
$p(r,\phi,s|r_0,\phi_0)=(1/D)\sum_m\exp[im(\phi-\phi_0)]p_m(r,s|r_0)$,
one notes that only the $m=\pm1$ terms contribute to $\langle x\rangle$ and $\langle y\rangle$. Furthermore, for $r=\sigma$, the square brackets simplify by virtue of the Wronskian of the modified Bessel functions, $K_m'(z)I_m(z)-K_m(z)I_m'(z)=1/z$, so that
\begin{equation}
  p_1(\sigma,s|r_0)=\frac1{q\sigma}\frac{K_1(qr_0)}{K_1'(q\sigma)}\,.
\end{equation}
Summarizing these expressions, one arrives at Eq.~\eqref{eq2dimavg}. The 3D result, Eq.~\eqref{eq3dimavg} is obtained along the same lines.

\end{appendix}


\bibliography{BD4HS}

\begin{thebibliography}{51}
\expandafter\ifx\csname natexlab\endcsname\relax\def\natexlab#1{#1}\fi
\expandafter\ifx\csname bibnamefont\endcsname\relax
  \def\bibnamefont#1{#1}\fi
\expandafter\ifx\csname bibfnamefont\endcsname\relax
  \def\bibfnamefont#1{#1}\fi
\expandafter\ifx\csname citenamefont\endcsname\relax
  \def\citenamefont#1{#1}\fi
\expandafter\ifx\csname url\endcsname\relax
  \def\url#1{\texttt{#1}}\fi
\expandafter\ifx\csname urlprefix\endcsname\relax\def\urlprefix{URL }\fi
\providecommand{\bibinfo}[2]{#2}
\providecommand{\eprint}[2][]{\url{#2}}

\bibitem[{\citenamefont{{Ermak}}(1975)}]{Ermak}
\bibinfo{author}{\bibfnamefont{D.~L.} \bibnamefont{{Ermak}}},
  \bibinfo{journal}{\jcp} \textbf{\bibinfo{volume}{62}}, \bibinfo{pages}{4189}
  (\bibinfo{year}{1975}).

\bibitem[{\citenamefont{Hansen and McDonald}(1989)}]{HansenMcDonaldBook}
\bibinfo{author}{\bibfnamefont{J.~P.} \bibnamefont{Hansen}} \bibnamefont{and}
  \bibinfo{author}{\bibfnamefont{I.~R.} \bibnamefont{McDonald}},
  \emph{\bibinfo{title}{Theory of Simple Liquid}} (\bibinfo{publisher}{Academic
  Press}, \bibinfo{address}{New York}, \bibinfo{year}{1989}),
  \bibinfo{edition}{2nd} ed.

\bibitem[{\citenamefont{Pusey}(1991)}]{PuseyChapter}
\bibinfo{author}{\bibfnamefont{P.~N.} \bibnamefont{Pusey}}, in
  \emph{\bibinfo{booktitle}{Liquids, Freezing and Glass Transition}}, edited by
  \bibinfo{editor}{\bibfnamefont{J.~P.} \bibnamefont{Hansen}},
  \bibinfo{editor}{\bibfnamefont{D.}~\bibnamefont{Levesque}}, \bibnamefont{and}
  \bibinfo{editor}{\bibfnamefont{J.}~\bibnamefont{Zinn-Justin}}
  (\bibinfo{publisher}{North-Holland}, \bibinfo{address}{Amsterdam},
  \bibinfo{year}{1991}), pp. \bibinfo{pages}{765--942}.

\bibitem[{\citenamefont{Cichocki and Hinsen}(1990)}]{Cichocki90}
\bibinfo{author}{\bibfnamefont{B.}~\bibnamefont{Cichocki}} \bibnamefont{and}
  \bibinfo{author}{\bibfnamefont{K.}~\bibnamefont{Hinsen}},
  \bibinfo{journal}{Physica A} \textbf{\bibinfo{volume}{166}},
  \bibinfo{pages}{473} (\bibinfo{year}{1990}).

\bibitem[{\citenamefont{Heyes and Melrose}(1993)}]{Heyes93}
\bibinfo{author}{\bibfnamefont{D.~M.} \bibnamefont{Heyes}} \bibnamefont{and}
  \bibinfo{author}{\bibfnamefont{J.~R.} \bibnamefont{Melrose}},
  \bibinfo{journal}{J.~Non-Newt.\ Fluid Mech.} \textbf{\bibinfo{volume}{46}},
  \bibinfo{pages}{1} (\bibinfo{year}{1993}).

\bibitem[{\citenamefont{Moriguchi et~al.}(1995)\citenamefont{Moriguchi,
  Kawasaki, and Kawakatsu}}]{Moriguchi95}
\bibinfo{author}{\bibfnamefont{I.}~\bibnamefont{Moriguchi}},
  \bibinfo{author}{\bibfnamefont{K.}~\bibnamefont{Kawasaki}}, \bibnamefont{and}
  \bibinfo{author}{\bibfnamefont{T.}~\bibnamefont{Kawakatsu}},
  \bibinfo{journal}{J.~Phys.\ II (France)} \textbf{\bibinfo{volume}{5}},
  \bibinfo{pages}{143} (\bibinfo{year}{1995}).

\bibitem[{\citenamefont{Schaertl and Sillescu}(1994)}]{Sillescu94}
\bibinfo{author}{\bibfnamefont{W.}~\bibnamefont{Schaertl}} \bibnamefont{and}
  \bibinfo{author}{\bibfnamefont{H.}~\bibnamefont{Sillescu}},
  \bibinfo{journal}{J.~Stat.~Phys.} \textbf{\bibinfo{volume}{74}},
  \bibinfo{pages}{687} (\bibinfo{year}{1994}).

\bibitem[{\citenamefont{{Foss} and {Brady}}(2000)}]{Foss2000}
\bibinfo{author}{\bibfnamefont{D.~R.} \bibnamefont{{Foss}}} \bibnamefont{and}
  \bibinfo{author}{\bibfnamefont{J.~F.} \bibnamefont{{Brady}}},
  \bibinfo{journal}{J.~Fluid Mech.} \textbf{\bibinfo{volume}{407}},
  \bibinfo{pages}{167} (\bibinfo{year}{2000}).

\bibitem[{\citenamefont{{Strating}}(1999)}]{Strating99}
\bibinfo{author}{\bibfnamefont{P.}~\bibnamefont{{Strating}}},
  \bibinfo{journal}{\pre} \textbf{\bibinfo{volume}{59}}, \bibinfo{pages}{2175}
  (\bibinfo{year}{1999}).

\bibitem[{\citenamefont{Tao et~al.}(2006)\citenamefont{Tao, den Otter, Dhont,
  and Briels}}]{Tao06}
\bibinfo{author}{\bibfnamefont{Y.-G.} \bibnamefont{Tao}},
  \bibinfo{author}{\bibfnamefont{W.~K.} \bibnamefont{den Otter}},
  \bibinfo{author}{\bibfnamefont{J.~K.~G.} \bibnamefont{Dhont}},
  \bibnamefont{and} \bibinfo{author}{\bibfnamefont{W.~J.}
  \bibnamefont{Briels}}, \bibinfo{journal}{J.~Chem.~Phys.}
  \textbf{\bibinfo{volume}{124}}, \bibinfo{pages}{134906}
  (\bibinfo{year}{2006}).

\bibitem[{\citenamefont{{Foffi} et~al.}(2005)\citenamefont{{Foffi},
  {De~Michele}, {Sciortino}, and {Tartaglia}}}]{Foffi05}
\bibinfo{author}{\bibfnamefont{G.}~\bibnamefont{{Foffi}}},
  \bibinfo{author}{\bibfnamefont{C.~D.} \bibnamefont{{De~Michele}}},
  \bibinfo{author}{\bibfnamefont{F.}~\bibnamefont{{Sciortino}}},
  \bibnamefont{and}
  \bibinfo{author}{\bibfnamefont{P.}~\bibnamefont{{Tartaglia}}},
  \bibinfo{journal}{Phys.~Rev.\ Lett.} \textbf{\bibinfo{volume}{94}},
  \bibinfo{pages}{078301} (\bibinfo{year}{2005}), \bibinfo{note}{where the
  algorithm used is incorrectly called Strating's}.

\bibitem[{\citenamefont{{de J.~Guevara-Rodr{\'{\i}}guez} and
  {Medina-Noyola}}(2003)}]{Guevara03}
\bibinfo{author}{\bibfnamefont{F.}~\bibnamefont{{de
  J.~Guevara-Rodr{\'{\i}}guez}}} \bibnamefont{and}
  \bibinfo{author}{\bibfnamefont{M.}~\bibnamefont{{Medina-Noyola}}},
  \bibinfo{journal}{\pre} \textbf{\bibinfo{volume}{68}},
  \bibinfo{pages}{011405} (\bibinfo{year}{2003}).

\bibitem[{Fuc()}]{FuchsPrivateComm}
\bibinfo{note}{Mathias Fuchs, private communication}.

\bibitem[{\citenamefont{{Brady}}(1993)}]{Brady93}
\bibinfo{author}{\bibfnamefont{J.~F.} \bibnamefont{{Brady}}},
  \bibinfo{journal}{\jcp} \textbf{\bibinfo{volume}{98}}, \bibinfo{pages}{3335}
  (\bibinfo{year}{1993}).

\bibitem[{\citenamefont{{Brady} and {Bossis}}(1988)}]{Brady88}
\bibinfo{author}{\bibfnamefont{J.~F.} \bibnamefont{{Brady}}} \bibnamefont{and}
  \bibinfo{author}{\bibfnamefont{G.}~\bibnamefont{{Bossis}}},
  \bibinfo{journal}{Annu.~Rev.~Fluid Mech.} \textbf{\bibinfo{volume}{20}},
  \bibinfo{pages}{111} (\bibinfo{year}{1988}).

\bibitem[{\citenamefont{Ladd}(1994)}]{Ladd94}
\bibinfo{author}{\bibfnamefont{A.~J.~C.} \bibnamefont{Ladd}},
  \bibinfo{journal}{J.~Fluid~Mech.} \textbf{\bibinfo{volume}{271}},
  \bibinfo{pages}{285} (\bibinfo{year}{1994}).

\bibitem[{\citenamefont{Ladd and Verberg}(2001)}]{Ladd01}
\bibinfo{author}{\bibfnamefont{A.~J.~C.} \bibnamefont{Ladd}} \bibnamefont{and}
  \bibinfo{author}{\bibfnamefont{R.}~\bibnamefont{Verberg}},
  \bibinfo{journal}{J.~Stat.~Phys.} \textbf{\bibinfo{volume}{104}},
  \bibinfo{pages}{1191} (\bibinfo{year}{2001}).

\bibitem[{\citenamefont{Cates et~al.}(2004)\citenamefont{Cates, Stratford,
  Adhikari, Stansell, Desplat, Pagonabarraga, and Wagner}}]{Rjoy04}
\bibinfo{author}{\bibfnamefont{M.~E.} \bibnamefont{Cates}},
  \bibinfo{author}{\bibfnamefont{K.}~\bibnamefont{Stratford}},
  \bibinfo{author}{\bibfnamefont{R.}~\bibnamefont{Adhikari}},
  \bibinfo{author}{\bibfnamefont{P.}~\bibnamefont{Stansell}},
  \bibinfo{author}{\bibfnamefont{J.-C.} \bibnamefont{Desplat}},
  \bibinfo{author}{\bibfnamefont{I.}~\bibnamefont{Pagonabarraga}},
  \bibnamefont{and} \bibinfo{author}{\bibfnamefont{A.~J.}
  \bibnamefont{Wagner}}, \bibinfo{journal}{J.~Phys.: Condens.~Matt.}
  \textbf{\bibinfo{volume}{16}}, \bibinfo{pages}{S3903} (\bibinfo{year}{2004}).

\bibitem[{\citenamefont{Adhikari et~al.}(2005)\citenamefont{Adhikari,
  Stratford, Cates, and Wagner}}]{Rjoy05}
\bibinfo{author}{\bibfnamefont{R.}~\bibnamefont{Adhikari}},
  \bibinfo{author}{\bibfnamefont{K.}~\bibnamefont{Stratford}},
  \bibinfo{author}{\bibfnamefont{M.~E.} \bibnamefont{Cates}}, \bibnamefont{and}
  \bibinfo{author}{\bibfnamefont{A.~J.} \bibnamefont{Wagner}},
  \bibinfo{journal}{Europhys.~Lett.} \textbf{\bibinfo{volume}{71}},
  \bibinfo{pages}{473} (\bibinfo{year}{2005}).

\bibitem[{\citenamefont{Hoogerbrugge and Koelman}(1992)}]{Hoogerbrugge92}
\bibinfo{author}{\bibfnamefont{P.~J.} \bibnamefont{Hoogerbrugge}}
  \bibnamefont{and} \bibinfo{author}{\bibfnamefont{J.~M.~V.~A.}
  \bibnamefont{Koelman}}, \bibinfo{journal}{Europhys.~Lett.}
  \textbf{\bibinfo{volume}{19}}, \bibinfo{pages}{155} (\bibinfo{year}{1992}).

\bibitem[{\citenamefont{Groot and Warren}(1997)}]{Groot97}
\bibinfo{author}{\bibfnamefont{R.~D.} \bibnamefont{Groot}} \bibnamefont{and}
  \bibinfo{author}{\bibfnamefont{P.~B.} \bibnamefont{Warren}},
  \bibinfo{journal}{J.~Chem.~Phys.} \textbf{\bibinfo{volume}{107}},
  \bibinfo{pages}{4423} (\bibinfo{year}{1997}).

\bibitem[{\citenamefont{Tanaka and Araki}(2000)}]{Tanaka00}
\bibinfo{author}{\bibfnamefont{H.}~\bibnamefont{Tanaka}} \bibnamefont{and}
  \bibinfo{author}{\bibfnamefont{T.}~\bibnamefont{Araki}},
  \bibinfo{journal}{Phys.~Rev.\ Lett.} \textbf{\bibinfo{volume}{85}},
  \bibinfo{pages}{1338} (\bibinfo{year}{2000}).

\bibitem[{\citenamefont{Tao et~al.}(2005)\citenamefont{Tao, den Otter, and
  Padding}}]{Tao05}
\bibinfo{author}{\bibfnamefont{Y.-G.} \bibnamefont{Tao}},
  \bibinfo{author}{\bibfnamefont{W.~K.} \bibnamefont{den Otter}},
  \bibnamefont{and} \bibinfo{author}{\bibfnamefont{J.~T.}
  \bibnamefont{Padding}}, \bibinfo{journal}{J. Chem. Phys.}
  \textbf{\bibinfo{volume}{122}}, \bibinfo{pages}{244903}
  (\bibinfo{year}{2005}).

\bibitem[{\citenamefont{Allen and Tildesley}(1987)}]{ATBook}
\bibinfo{author}{\bibfnamefont{M.~P.} \bibnamefont{Allen}} \bibnamefont{and}
  \bibinfo{author}{\bibfnamefont{D.~J.} \bibnamefont{Tildesley}},
  \emph{\bibinfo{title}{Computer Simulation of Liquids}}
  (\bibinfo{publisher}{Clarendon Press, Oxford}, \bibinfo{year}{1987}),
  \bibinfo{edition}{2nd} ed.

\bibitem[{\citenamefont{Kloeden and Platen}(1999)}]{KloedenPlatenBook}
\bibinfo{author}{\bibfnamefont{P.~E.} \bibnamefont{Kloeden}} \bibnamefont{and}
  \bibinfo{author}{\bibfnamefont{E.}~\bibnamefont{Platen}},
  \emph{\bibinfo{title}{Numerical solution of stochastic differential
  equations}}, vol.~\bibinfo{volume}{23} of \emph{\bibinfo{series}{Applications
  of Mathematics}} (\bibinfo{publisher}{Springer}, \bibinfo{year}{1999}),
  \bibinfo{edition}{3rd} ed.

\bibitem[{\citenamefont{Xue and Grest}(1990)}]{Xue90}
\bibinfo{author}{\bibfnamefont{W.}~\bibnamefont{Xue}} \bibnamefont{and}
  \bibinfo{author}{\bibfnamefont{G.~S.} \bibnamefont{Grest}},
  \bibinfo{journal}{\prl} \textbf{\bibinfo{volume}{64}}, \bibinfo{pages}{419}
  (\bibinfo{year}{1990}).

\bibitem[{\citenamefont{Turq et~al.}(1977)\citenamefont{Turq, Lantelme, and
  Friedman}}]{Turq77}
\bibinfo{author}{\bibfnamefont{P.}~\bibnamefont{Turq}},
  \bibinfo{author}{\bibfnamefont{F.}~\bibnamefont{Lantelme}}, \bibnamefont{and}
  \bibinfo{author}{\bibfnamefont{L.}~\bibnamefont{Friedman}},
  \bibinfo{journal}{\jcp} \textbf{\bibinfo{volume}{66}}, \bibinfo{pages}{3039}
  (\bibinfo{year}{1977}).

\bibitem[{\citenamefont{Helfand}(1978)}]{Helfand78}
\bibinfo{author}{\bibfnamefont{E.}~\bibnamefont{Helfand}},
  \bibinfo{journal}{\jcp} \textbf{\bibinfo{volume}{69}}, \bibinfo{pages}{1010}
  (\bibinfo{year}{1978}).

\bibitem[{\citenamefont{Iniesta and {Garc{\'\i}a de la
  Torre}}(1990)}]{Iniesta90}
\bibinfo{author}{\bibfnamefont{A.}~\bibnamefont{Iniesta}} \bibnamefont{and}
  \bibinfo{author}{\bibfnamefont{J.}~\bibnamefont{{Garc{\'\i}a de la Torre}}},
  \bibinfo{journal}{\jcp} \textbf{\bibinfo{volume}{92}}, \bibinfo{pages}{2015}
  (\bibinfo{year}{1990}).

\bibitem[{\citenamefont{Honeycutt}(1992)}]{Honeycutt92}
\bibinfo{author}{\bibfnamefont{R.}~\bibnamefont{Honeycutt}},
  \bibinfo{journal}{\pra} \textbf{\bibinfo{volume}{45}}, \bibinfo{pages}{600}
  (\bibinfo{year}{1992}).

\bibitem[{\citenamefont{{Bra\'nka} and Heyes}(1998)}]{Branka98}
\bibinfo{author}{\bibfnamefont{A.~C.} \bibnamefont{{Bra\'nka}}}
  \bibnamefont{and} \bibinfo{author}{\bibfnamefont{D.~M.} \bibnamefont{Heyes}},
  \bibinfo{journal}{\pre} \textbf{\bibinfo{volume}{58}}, \bibinfo{pages}{2611}
  (\bibinfo{year}{1998}).

\bibitem[{\citenamefont{{Skeel} and Izaguirre}(2002)}]{Skeel02}
\bibinfo{author}{\bibfnamefont{R.~D.} \bibnamefont{{Skeel}}} \bibnamefont{and}
  \bibinfo{author}{\bibfnamefont{J.~A.} \bibnamefont{Izaguirre}},
  \bibinfo{journal}{Molec.~Phys.} \textbf{\bibinfo{volume}{100}},
  \bibinfo{pages}{3885} (\bibinfo{year}{2002}).

\bibitem[{\citenamefont{Wang and Skeel}(2003)}]{Wang03}
\bibinfo{author}{\bibfnamefont{W.}~\bibnamefont{Wang}} \bibnamefont{and}
  \bibinfo{author}{\bibfnamefont{R.~D.} \bibnamefont{Skeel}},
  \bibinfo{journal}{Molec.~Phys.} \textbf{\bibinfo{volume}{101}},
  \bibinfo{pages}{2149} (\bibinfo{year}{2003}).

\bibitem[{\citenamefont{Ricci and Ciccotti}(2003)}]{Ricci03}
\bibinfo{author}{\bibfnamefont{A.}~\bibnamefont{Ricci}} \bibnamefont{and}
  \bibinfo{author}{\bibfnamefont{G.}~\bibnamefont{Ciccotti}},
  \bibinfo{journal}{Molec.~Phys.} \textbf{\bibinfo{volume}{101}},
  \bibinfo{pages}{1927} (\bibinfo{year}{2003}).

\bibitem[{\citenamefont{Kannan and Lakshmikantham}(2002)}]{KannanBook}
\bibinfo{author}{\bibfnamefont{D.}~\bibnamefont{Kannan}} \bibnamefont{and}
  \bibinfo{author}{\bibfnamefont{V.}~\bibnamefont{Lakshmikantham}},
  \emph{\bibinfo{title}{Handbook of stochastic analysis and applications}}
  (\bibinfo{publisher}{Marcel Dekker}, \bibinfo{year}{2002}).

\bibitem[{\citenamefont{{Rapaport}}(2004)}]{RapaBook}
\bibinfo{author}{\bibfnamefont{D.~C.} \bibnamefont{{Rapaport}}},
  \emph{\bibinfo{title}{{The Art of Molecular Dynamics Simulation}}}
  (\bibinfo{publisher}{Cambridge University Press}, \bibinfo{year}{2004}).

\bibitem[{\citenamefont{Allen et~al.}(1989)\citenamefont{Allen, Frenkel, and
  Talbot}}]{Allen89}
\bibinfo{author}{\bibfnamefont{M.}~\bibnamefont{Allen}},
  \bibinfo{author}{\bibfnamefont{D.}~\bibnamefont{Frenkel}}, \bibnamefont{and}
  \bibinfo{author}{\bibfnamefont{J.}~\bibnamefont{Talbot}},
  \bibinfo{journal}{Comput.\ Phys.\ Rep.} \textbf{\bibinfo{volume}{9}},
  \bibinfo{pages}{301} (\bibinfo{year}{1989}).

\bibitem[{\citenamefont{Luding et~al.}(1994)\citenamefont{Luding, Herrmann, and
  Blumen}}]{Luding94}
\bibinfo{author}{\bibfnamefont{S.}~\bibnamefont{Luding}},
  \bibinfo{author}{\bibfnamefont{H.~J.} \bibnamefont{Herrmann}},
  \bibnamefont{and} \bibinfo{author}{\bibfnamefont{A.}~\bibnamefont{Blumen}},
  \bibinfo{journal}{Phys.~Rev.~E} \textbf{\bibinfo{volume}{50}},
  \bibinfo{pages}{3100} (\bibinfo{year}{1994}).

\bibitem[{\citenamefont{Luding}(1995)}]{Luding95}
\bibinfo{author}{\bibfnamefont{S.}~\bibnamefont{Luding}},
  \bibinfo{journal}{Phys.~Rev.~E} \textbf{\bibinfo{volume}{52}},
  \bibinfo{pages}{4442} (\bibinfo{year}{1995}).

\bibitem[{\citenamefont{{Williams} and {Mackintosh}}(1996)}]{Williams96}
\bibinfo{author}{\bibfnamefont{D.~R.~M.} \bibnamefont{{Williams}}}
  \bibnamefont{and} \bibinfo{author}{\bibfnamefont{F.~C.}
  \bibnamefont{{Mackintosh}}}, \bibinfo{journal}{\pre}
  \textbf{\bibinfo{volume}{54}}, \bibinfo{pages}{9} (\bibinfo{year}{1996}).

\bibitem[{\citenamefont{{Peng} and {Ohta}}(1998)}]{Peng98}
\bibinfo{author}{\bibfnamefont{G.}~\bibnamefont{{Peng}}} \bibnamefont{and}
  \bibinfo{author}{\bibfnamefont{T.}~\bibnamefont{{Ohta}}},
  \bibinfo{journal}{\pre} \textbf{\bibinfo{volume}{58}}, \bibinfo{pages}{4737}
  (\bibinfo{year}{1998}).

\bibitem[{\citenamefont{{van Noije} et~al.}(1999)\citenamefont{{van Noije},
  {Ernst}, {Trizac}, and {Pagonabarraga}}}]{VanNoije99}
\bibinfo{author}{\bibfnamefont{T.~P.~C.} \bibnamefont{{van Noije}}},
  \bibinfo{author}{\bibfnamefont{M.~H.} \bibnamefont{{Ernst}}},
  \bibinfo{author}{\bibfnamefont{E.}~\bibnamefont{{Trizac}}}, \bibnamefont{and}
  \bibinfo{author}{\bibfnamefont{I.}~\bibnamefont{{Pagonabarraga}}},
  \bibinfo{journal}{\pre} \textbf{\bibinfo{volume}{59}}, \bibinfo{pages}{4326}
  (\bibinfo{year}{1999}).

\bibitem[{\citenamefont{{Hanna} et~al.}(1982)\citenamefont{{Hanna}, {Hess}, and
  {Klein}}}]{Hanna82}
\bibinfo{author}{\bibfnamefont{S.}~\bibnamefont{{Hanna}}},
  \bibinfo{author}{\bibfnamefont{W.}~\bibnamefont{{Hess}}}, \bibnamefont{and}
  \bibinfo{author}{\bibfnamefont{R.}~\bibnamefont{{Klein}}},
  \bibinfo{journal}{Physica A} \textbf{\bibinfo{volume}{111}},
  \bibinfo{pages}{181} (\bibinfo{year}{1982}).

\bibitem[{\citenamefont{Ackerson and Fleishman}(1982)}]{Ackerson82}
\bibinfo{author}{\bibfnamefont{B.~J.} \bibnamefont{Ackerson}} \bibnamefont{and}
  \bibinfo{author}{\bibfnamefont{L.}~\bibnamefont{Fleishman}},
  \bibinfo{journal}{Journal of Chemical Physics} \textbf{\bibinfo{volume}{76}},
  \bibinfo{pages}{2675} (\bibinfo{year}{1982}).

\bibitem[{\citenamefont{{Polyanin}}(2002)}]{PolyaninBook}
\bibinfo{author}{\bibfnamefont{A.~D.} \bibnamefont{{Polyanin}}},
  \emph{\bibinfo{title}{{Handbook of linear partial differential equations for
  engineers and scientists }}} (\bibinfo{publisher}{CRC Press},
  \bibinfo{year}{2002}).

\bibitem[{\citenamefont{Foister and van~de Ven}(1980)}]{Foister80}
\bibinfo{author}{\bibfnamefont{R.~T.} \bibnamefont{Foister}} \bibnamefont{and}
  \bibinfo{author}{\bibfnamefont{T.~G.~M.} \bibnamefont{van~de Ven}},
  \bibinfo{journal}{J.~Fluid.\ Mech.} \textbf{\bibinfo{volume}{96}},
  \bibinfo{pages}{105} (\bibinfo{year}{1980}).

\bibitem[{\citenamefont{Courant and Hilbert}(1953)}]{CourantHilbertBook}
\bibinfo{author}{\bibfnamefont{R.}~\bibnamefont{Courant}} \bibnamefont{and}
  \bibinfo{author}{\bibfnamefont{D.}~\bibnamefont{Hilbert}},
  \emph{\bibinfo{title}{Methods of Mathematical Physics}},
  vol.~\bibinfo{volume}{I} (\bibinfo{publisher}{Wiley \& Sons},
  \bibinfo{address}{New York}, \bibinfo{year}{1953}).

\bibitem[{\citenamefont{Valk\'o and Abate}(2004)}]{LTinversion}
\bibinfo{author}{\bibfnamefont{P.~P.} \bibnamefont{Valk\'o}} \bibnamefont{and}
  \bibinfo{author}{\bibfnamefont{J.}~\bibnamefont{Abate}},
  \bibinfo{journal}{Comput.~Math.~Appl.} \textbf{\bibinfo{volume}{48}},
  \bibinfo{pages}{629} (\bibinfo{year}{2004}),
  \bibinfo{note}{\textit{Mathematica} source code from
  \texttt{http://library.wolfram.com/infocenter/MathSource/4738/}}.

\bibitem[{\citenamefont{Heyes and Bra\'nka}(1998)}]{Heyes98}
\bibinfo{author}{\bibfnamefont{D.~M.} \bibnamefont{Heyes}} \bibnamefont{and}
  \bibinfo{author}{\bibfnamefont{A.~C.} \bibnamefont{Bra\'nka}},
  \bibinfo{journal}{Molec.~Phys.} \textbf{\bibinfo{volume}{94}},
  \bibinfo{pages}{447} (\bibinfo{year}{1998}).

\bibitem[{\citenamefont{Crank}(1975)}]{CrankBook}
\bibinfo{author}{\bibfnamefont{J.}~\bibnamefont{Crank}},
  \emph{\bibinfo{title}{The mathematics of diffusion}}
  (\bibinfo{publisher}{Clarendon Press, Oxford}, \bibinfo{year}{1975}),
  \bibinfo{edition}{2nd} ed.

\bibitem[{\citenamefont{Schulten and Kosztin}(2000)}]{SchultenEbook}
\bibinfo{author}{\bibfnamefont{K.}~\bibnamefont{Schulten}} \bibnamefont{and}
  \bibinfo{author}{\bibfnamefont{I.}~\bibnamefont{Kosztin}},
  \emph{\bibinfo{title}{Lectures in theoretical biophysics}}
  (\bibinfo{year}{2000}),
  \urlprefix\url{http://www.caam.rice.edu/~cox/stoch/lectheobio.pdf}.

\end{thebibliography}


\end{document}